\documentclass[11pt]{article}
\usepackage[margin=1in]{geometry}
\usepackage{lmodern}
\usepackage{sidecap}
\usepackage{xspace}
\usepackage{doi}
\usepackage{hyperref} % Should come last
\usepackage{tikz}
\usepackage{pgfgantt}
\usepackage{graphicx,pbox}
\usepackage{xcolor,hyperref,colortbl}
\usepackage{wrapfig}
\usepackage{enumitem}
\usepackage{gensymb}
\usepackage{comment}
\newcommand\pubnumber{}
\newcommand\pubdate{\today}
\newcommand\pubblock{\rightline{\begin{tabular}{l} \pubnumber\\
          \pubdate \end{tabular}}}
\newcommand{\beq}{\begin{equation}}
\newcommand{\eeq}{\end{equation}}
\newcommand{\exclude}[1]{}

\def\Title#1{\begin{center} {\Large #1 } \end{center}}
\def\Author#1{\begin{center}{ \sc #1} \end{center}}

\newcommand\snowmass{\begin{center}\rule[-0.2in]{\hsize}{0.01in}\\\rule{\hsize}{0.01in}\\
\vskip 0.1in Submitted to the  Proceedings of the US Community Study\\ 
on the Future of Particle Physics (Snowmass 2021)\\ 
\rule{\hsize}{0.01in}\\\rule[+0.2in]{\hsize}{0.01in} \end{center}}
\begin{document}
\begin{titlepage}
\snowmass
\pubblock
 
%\vfill
\Title{Probing Fundamental Physics With Multi-Modal Cosmic Ray  Events}

\Author{D.~Beznosko$^{1}$, K.~Baigarin$^{2}$, R.~Beisembaev $^{3}$, E.~Beisembaeva$^{3}$, E.~Gladysz-Dziadu$^{4}$\footnote{Retired}, V.~Ryabov$^{3}$, T.~Sadykov$^{5}$, S.~Shaulov$^{3}$,  V.~Shiltsev$^{6}$, A.~Stepanov$^{3}$, M.~Vildanova$^{3}$, A.~Zhitnitsky$^{7}$, V.~Zhukov$^{3}$}

\noindent
$^{1}${ Clayton State University, Morrow, GA 30260, USA}\\
$^{2}${ Nazarbayev University, Nur-Sultan,  010000, Kazakhstan} \\
$^{3}${ P.N.Lebedev Physical Institute of the Russian Academy of Sciences, Moscow, 119991, Russia} \\
$^{4}${ Institute of Nuclear Physics, Polish Academy of Sciences, Cracow, 31-342, Poland}\\
$^{5}${ Satbayev University, Institute of Physics and Technology, Almaty,  050000, Kazakhstan} \\
$^{6}${ Fermi National Accelerator Laboratory, Batavia, IL 60510, USA} \\
$^{7}${ University of British Columbia, Vancouver, V6T 1Z1, BC, Canada}\\

%\maketitle

%\date{April 2022}

%\maketitle
\end{titlepage}
\newpage

%\tableofcontents
\newpage

\addcontentsline{toc}{section}{Executive Summary}
\section*{Executive Summary}
Since the middle of last century, researchers have observed EAS events with more than one pulse from the passage of particles in each detector. These events were called events with a delayed particle at the time, now they are also called multi-modal events. The first registration of such events was described by J.~Jelley and W.~Whitehouse \cite{Jelley_1953} in 1953. Later, EAS exhibiting the unusual time structures were studied by several independent experiments, see e.g. \cite{PhysRev.128.2384}. EAS events with delayed particles were studied by British and US research groups in the 1960s-1980s, since the 1970s, delayed particles in EAS have been studied in Japan  and also carried out at Moscow State University (MSU) and at LPI (TSHASS) and other groups. The conclusions that these articles give do not match and point to the lack of understanding of the nature of events with delayed particles.

The {\it Horizon-T} experiment in Tien Shan is based on the idea of measuring the time at which EAS disc passes the observation level with nanosecond accuracy. The detector system \cite{beisembaev2019performance} consists of ten charged particles registration points located at distances of up to several hundred meters from each other. The points are equipped with detectors based on registration of Cherenkov radiation in glass and registration of scintillation light in polystyrene. The detectors register the arrival times of charged particles at the observation level with a resolution of $\sim$2 ns, as required to study the spatial and temporal characteristics of the EAS and the structure of the multi-modal events specifically. Over the period of the Horizon-T data taking since 2017 to present, a large number of multi-modal events were detected. The data has presented numerous challenges that show the direction towards the further development of the detector system and of the analysis methods and techniques that could be applied to these multi-modal events \cite{2019EPJWC.20806002B}.

These recently observed Multi-Modal Cosmic Rays Events (MME) containing multiple peaks separated by tens to hundreds of ns are one of the most puzzling phenomena in high energy cosmic rays. These unusual cosmic ray events offer a potential new insight into the ultra-high energy astrophysics, the physics of fundamental particle interactions and cosmology. Surely, {\bf further experimental studies and independent verification of these events both at lower and high altitudes are needed to better understand the MME phenomenology}. Theoretical insights are very much needed as well to guide future studies. It was recently proposed in \cite{Zhitnitsky:2021qhj} that these MMEs might be result of the dark matter annihilation events within the so-called axion quark nugget (AQN) dark matter model, which was originally invented for completely different purpose to explain the observed similarity between the dark and the visible components in the Universe, i.e. $\Omega_{DM} \approx \Omega_{visible}$, without any fitting parameters. 

Earlier, in the HADRON experiment on Tien Shan, the possible presence of a non-nuclear component in primary cosmic rays was established and it was assumed that this component could consist of stable particles of strange quark matter, strangelets \cite{Shaulov:1996zy} \cite{Shaulov:2021xld} \cite{Shepetov:2021vnr}. So with a high probability, this version of the interpretation of the MME  is preferable. Such a hypothesis was proposed in \cite{gladysz2021multimodal}, where it was hypothesised that experimental observations of the MME could be the manifestation of penetration of a strangelets through the matter. {\bf We strongly encourage further discussions and elaborations on these and other physics explanations of the unusual  multi-modal cosmic rays events.} 

We would like to attract attention to quite interdisciplinary nature of our proposal as it touches and overlaps with many fields of present-day active studies, including DM, baryogenesis, CR, nuclear physics, AMO, particle physics, astrophysics, cosmology, etc. Given limited success of conventional approaches to DM candidates such as WIMPs over the past few decades (as the LHC and dozen of DM detectors were no successful so far), we believe the proposed experimental studies of the MMEs together with further theoretical developments towards understanding their nature might help to formulate the so-much-required paradigm shift. 

\newpage

\section{Introduction: Multi-Modal Events Observations To Date} 

An Extensive Air Shower (EAS) is a cascade of secondary particles in the atmosphere formed as a result of the interaction of very high energy 10$^{14-20}$ eV primary cosmic particles with atomic nuclei in the Earth's atmosphere \cite{rao1998extensive}. The secondaries registered at the observation level come of several main components: electron-photon, hadron, muon, as well as Vavilov-Cherenkov radiation.  Ultra-relativistic particles that are generated in the EAS cascade processes arrive at the observation level in a shape of a disc with up to several kilometers in diameter depending on the energy of the primary cosmic ray (PCR). Thickness of the disc can vary from about a meter (few nanoseconds in time) near the axis of the EAS to hundreds of meters at its periphery \cite{engel2011extensive}. The most common type of EAS detectors system is a network of individual detectors that occupies a certain area large enough to encompass the EAS disk at the observation level as well as to register a sufficient number of events. Most often, the analysis uses the following data coming from individual detectors: the arrival time of the disc to determine the directions of PCR, and the particle density to estimate PCR energy and determine the position of the EAS axis relative to the detector system.

    \begin{figure}[ht]
	\centering
	%\captionsetup{justification=raggedright}
	\includegraphics[width=0.49\linewidth]{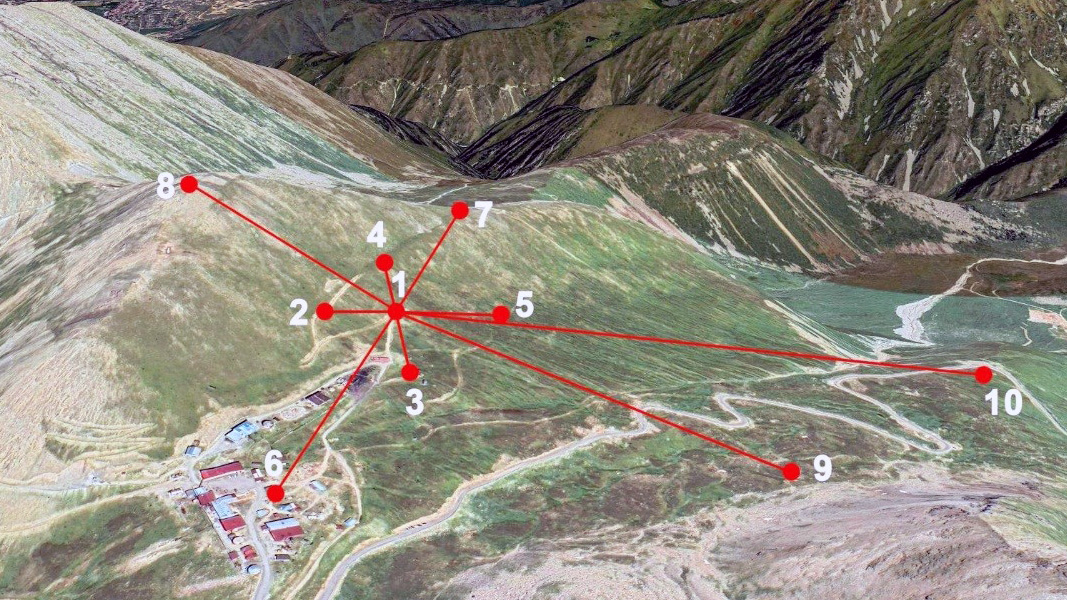}
	\includegraphics[width=0.49\linewidth]{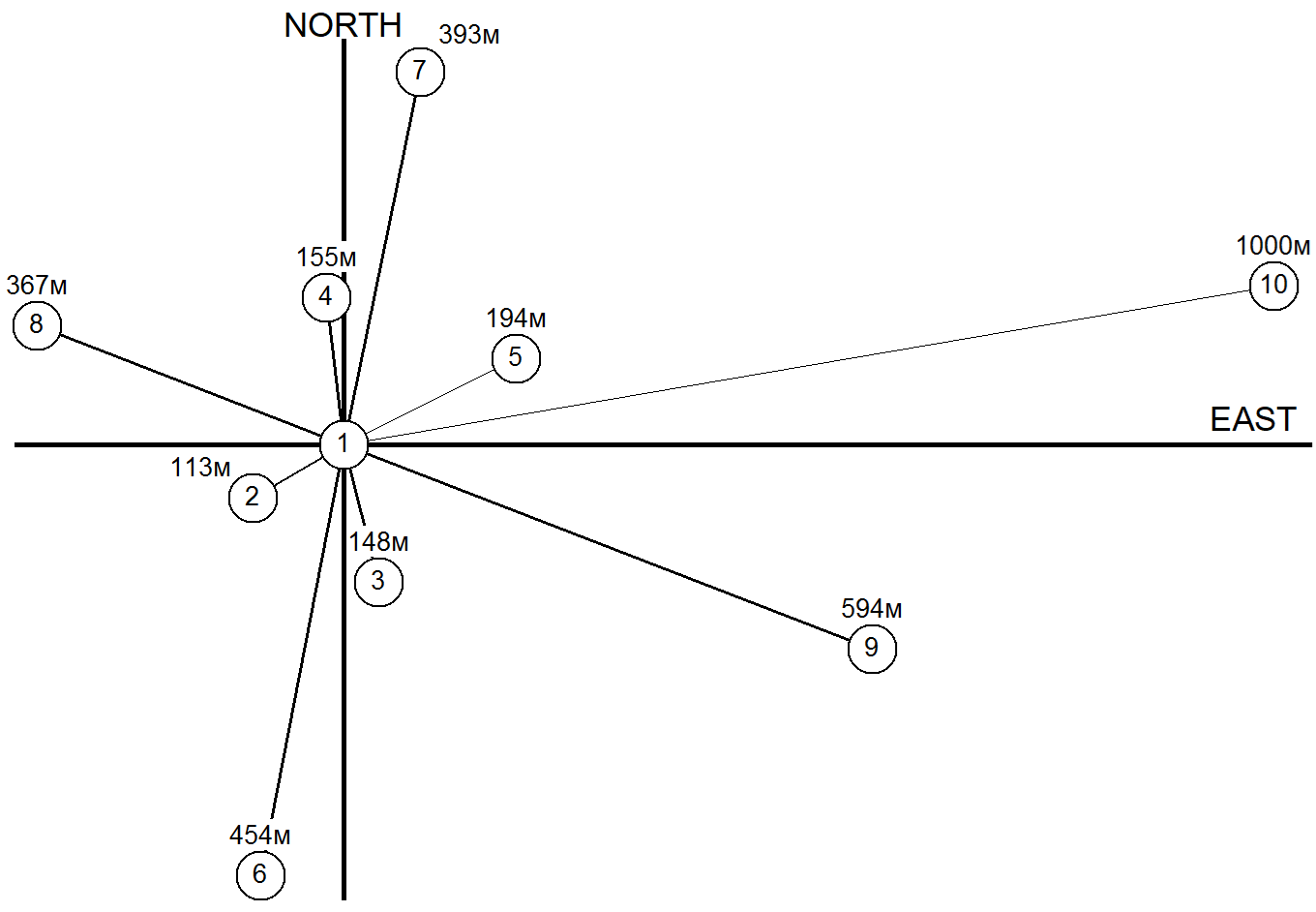}
	\caption{Arrangement of ten detection points in the Horizon-T array. Indicated are distances from the detection points to the array center at point 1. Aerial view and geometry of the Horizon-T instrument with the location of 10 detectors.}.  
	\label{Layout}
\end{figure}

The first observations of EAS events with more than one pulse from the passage of particles in each detector -- then called {\it events with  delayed particles} - were reported by J.~Jelley and W.~Whitehouse in 1953 \cite{Jelley_1953}. Then the EAS events with delayed particles were studied by British and US research groups in the 1960s-1980s \cite{linsley1962arrival, mincer1985search}, later joined by other countries \cite{kawamoto1985search, bhat1982delayed, agnetta1997time}, including several experiments in Russia \cite{glushkov1998temporal, budnev2009tunka, atrashkevich1997temporal, garipov2020search}. These experiments employed different detectors, were of various spatial scales and located at different altitudes. The Horizon-T detector system at Tien Shan \cite{beisembaev2009delayed} is designed and built with the fastest (~2 ns digitization resolution) particle detectors and with the scale of $>$ km as briefly described below. In all Horizon-T physics runs the unusual events with delayed pulses with several maxima have been recorded. These events cannot be fully explained by the physical processes of the standard model of EAS.

In Ref.\cite{Zhitnitsky:2021qhj} we summarize the most puzzling features of 
recent Horizon-1T observations  \cite{Beisembaev:2016cyg,2017EPJWC.14514001B,2019EPJWC.20806002B,Beznosko:2019cI,Beisembaev:2019nzd}: 
  
  {\it {\bf 1}. ``clustering puzzle":} Two or more peaks separated by $\sim 10^2$   ns are present in several  detection  points, while entire event may last $\sim 10^3$ ns. It can be viewed as many fronts separated by $\sim (10^2-10^3)$   ns, instead of a single  front; 
  
   {\it {\bf 2}. ``particle density puzzle":} The number density of particles recorded at different  detection  points apparently weakly  dependent  on 
   distance from  Extensive Air Showers (EAS) axis;
   
      {\it {\bf 3}. ``pulse width puzzle":} The width of each individual pulse is around $(20-35)$ ns and apparently does not   depend  on  distance from    EAS axis;

   {\it {\bf 4}. ``intensity puzzle":} The observed intensity of the events (measured in units of a number of particles per unit area) is of order $\rho\sim (100-300) \rm m^{-2}$
   when measured at distances $(300-800)$ m from EAS axis. Such intensity would   correspond to   the CR energy of the primary particle on the level $E_{p}> 10^{19}$ eV which would have  dramatically different event rate in comparison with observed rate recorded by Horizon-T. 

\subsection{Horizon-T experiment}

The Horizon-T experiment is deployed at the  LPI RAS Tien Shan High-Altitude Scientific Station located at an altitude of 3340 meters above sea level near the city of Almaty (Republic of Kazakhstan). The experiment is aimed at studying the temporal structure of extensive air showers from primary cosmic rays with energies above 10$^{16}$ eV, as well as studying delayed particles from the main shower front. 
Horizon-T comprises 10 detection points spaced up to 1400 meters from each other and tied to a common center - see Fig.\ref{Layout}. 
The points are equipped with detectors based on registration of Cherenkov radiation in glass and registration of scintillation light in polystyrene. The detectors record the arrival times of charged particles at the observation level with a resolution of $O$(2 ns), as required to study the spatial and temporal characteristics of the EAS. 
The EAS-induced signals from all the detectors are transmitted via high-frequency cables to the central point and digitized with a resolution of 2 ns.

    \begin{figure}[ht]
	\centering
	%\captionsetup{justification=raggedright}
	\includegraphics[width=0.85\linewidth]{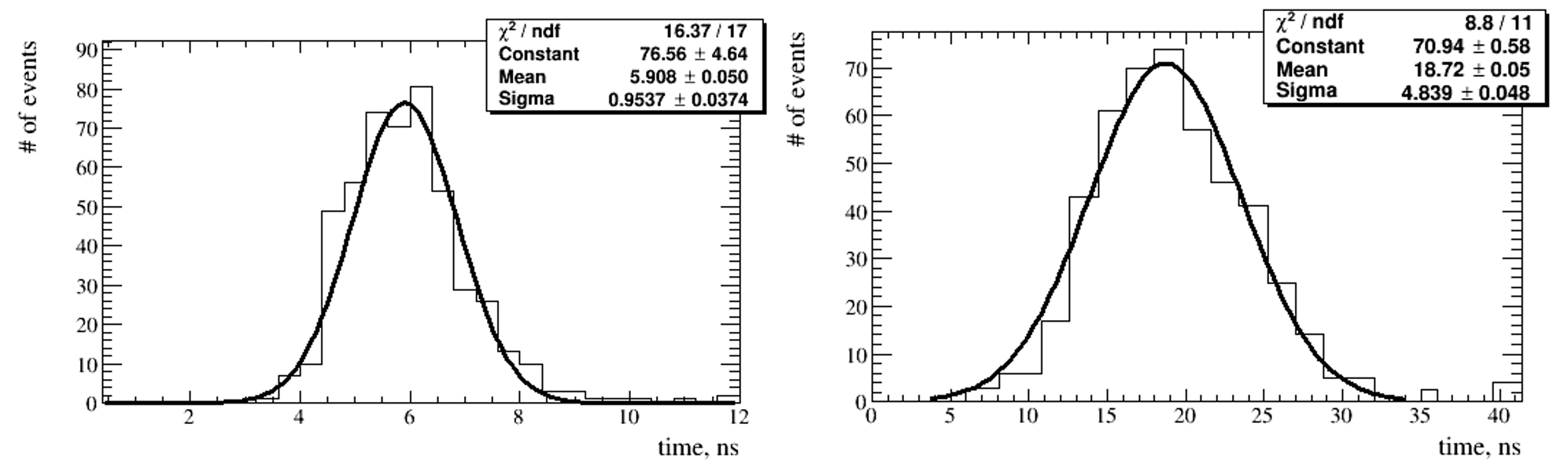}
	\caption{Pulse rise time (left) and total pulse width (right) of the Horizon-T scintillation detectors with Hamamatsu R7723 PMTs.}.  
	\label{pulses}
\end{figure}

The experiment employs scintillation radiation detectors (SC-detectors) with a particle detection medium of area of 100$\times$100 cm$^2$ and a thickness of 10 cm (5 cm in points 6, 7, 9, 10) oriented in the XY plane (horizontal). The detection medium used is P-terphenyl in polystyrene. The signal is read by Hamamatsu R7723 photomultipliers (PMT) with a diameter of 51 mm (in the assembly R7723 Y003) in points 1-8, and Hamamatsu R1250 with a diameter of 127 mm (in the assembly H6527) in points 9 and 10. Points 1-5 are additionally equipped with fast detectors of Cherenkov radiation in glass (GL-detectors) with the detection medium area of 50$\times$50 cm$^2$ and a thickness of 3 cm, oriented in the XY plane (horizontal) with Hamamatsu R7723 PMT readout. The pulse rise time for glass detectors is ~2.1 ns and total pulse width is ~ 5 ns.  Each detector is connected by a high-frequency cable to a data acquisition system located in the detection point 1. 

The Horizon-T system has the EAS detection thresholds energy varying from 10$^{16}$ eV at zero zenith angle to 2$\cdot 10^{18}$eV at 80 degrees. The EAS pulse width uncertainty is several ns and depends on  the scintillator light yield time ($\sim$5 ns in scintillator or $\sim$0.5 ns in glass), PMT electron transit time spread ($\sim$1.2 ns), ADC bin width (2 ns), and some pulse broadening in the high-frequency signal cable depending on the cable length - see Fig.\ref{pulses}.   

That  considerable recent improvement in time resolution (to the level of a few  ns) and the DAQ system able to recognize signals with complicated (multi-pulse) time structure have allowed the  Horizon-T collaboration  dramatically improve the detection, collection efficiency and analysis of the MMEs. In particular, with $\sim 4000$ hours of operation in 2016-2019 several years, about 30 thousands high-energy CR events were recorded out of which more than a hundred of multi-modal events with pulse modes separated by hundreds of ns. In several events, the time separation reached up to a $\mu$s. Recently, 26940 events with energy in excess of 2$\cdot 10^{16}$ eV were recorded over 1896 hours of operation in the period from December 12, 2019 to March 19, 2020. Out of those, 962 events with energies above $10^{17}$ eV had an unusual time structure including 217 impulses with two maxima. The number of EASs with a delay of the second pulse relative to the first pulse of less than 100 ns was 37, while 33 had the delay of more than 1000 ns.
  
      \begin{figure}[ht]
	\centering
	%\captionsetup{justification=raggedright}
	\includegraphics[width=0.85\linewidth]{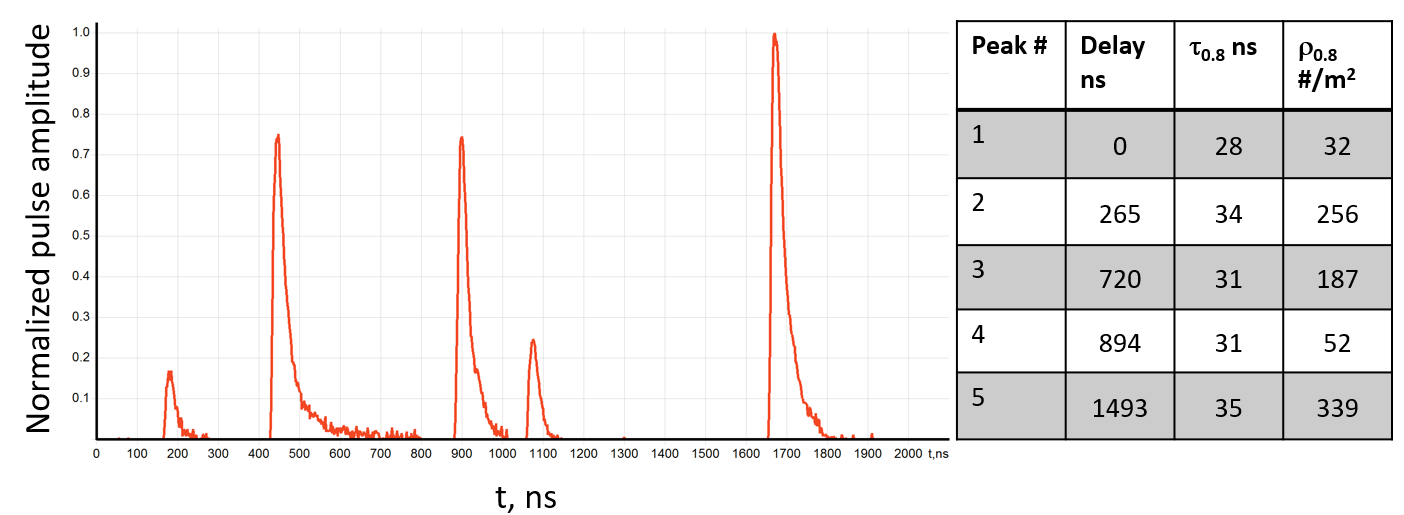}
	\caption{A typical MME event recorded  on April 6, 2018 by Horizon-T detector at point $\#9$, adopted from \cite{Beznosko:2019cI}. All pulses are recorded at a single detection point. Delay times,  the width of each peak $\tau_{0.8}$ in ns, and the particle density $\rho_{0.8}$  per  $\rm m^{-2}$ within $\tau_{0.8}$ are shown in the table. }.  
	\label{MME}
\end{figure}

 \subsection{Multi-modal events observations}
 \label{confronts}
 
%    \begin{figure}[ht]
%	\centering
%	%\captionsetup{justification=raggedright}
%	\includegraphics[width=0.85\linewidth]{1.pdf}
%	\caption{top: Aerial view and geometry of the Horizon-T instrument   with location of 10 detectors, adopted from \cite{Beznosko:2019cI};
%	bottom: A typical MME event recorded  on April 6.2018 by the Horizon-T instrument at point $\#9$, adopted from \cite{Beznosko:2019cI}. All pulses are recorded at a single detection point. Delay times,  the width of each peak $\tau_{0.8}$ in ns, and the particle density $\rho_{0.8}$  per  $\rm m^{-2}$ within $\tau_{0.8}$ are also shown in the table. }.  
%	\label{pulses}
%\end{figure}

 \begin{figure}[ht]
	\centering
	%\captionsetup{justification=raggedright}
	\includegraphics[width=0.89\linewidth]{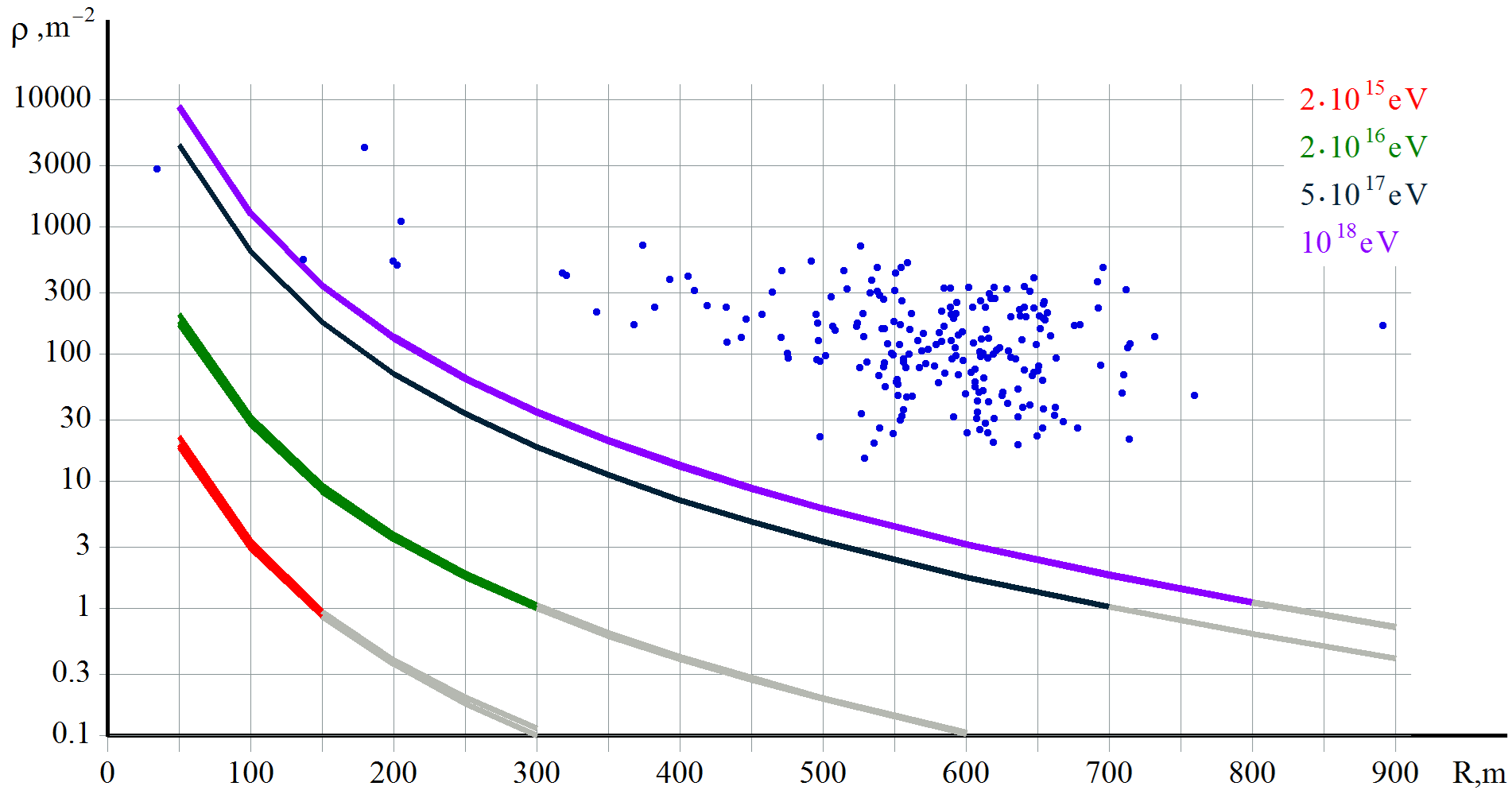}
	\caption{ Solid lines: the particle density distribution $\rho(R)$ in simulated EAS disk versus  distance from axis for different energies shown by different colours, depending on energy of the CR. Blue dots: cumulative particle density for each bimodal pulse vs. distance to EAS axis, adopted from \cite{Beznosko:2019cI}.}.  
	\label{density}
\end{figure}

 Here we briefly describe the MME observations and argue why the observed  events are inconsistent with conventional CR interpretation. In the conventional picture of the EAS, it is normally assumed that a) the EAS can be thought as a pancake-like disk with well defined EAS axis; b) the EAS represents an uniform structure without any breaks; c) the particle density drops smoothly with distance when moving away from the core, while d) the thickness of the EAS pancake  increases with the distance from the core. This conventional picture describes well an overwhelming majority of the cosmic ray events but is in dramatic contradiction with  observations of rare but well-detectable MME events.
 
{\bf 1.}  Indeed, a typical MME is shown in Fig. \ref{MME} where a complicated temporal features are explicitly seen. Several peaks separated by $\sim 10^2$   ns in a single detector represent  the {\it ``clustering puzzle"}, listed  above. In the conventional EAS   picture one should see a single pulse in each given detector with the amplitude which   depends on the distance from the EAS axis. It is not what actually observed by the Horizon-T.
   
{\bf 2.} Fig. \ref{density} manifests the {\it ``particle density puzzle"}: particle density distribution $\rho(R)$ for different CR energies in the EAS disk versus  distance from axis are simulated with CORSIKA code \cite{heck1998corsika} in shown by solid lines in different colors. In particular, for energy of the primary particle on the level of $10^{17}$ eV one should expect  a strong  suppression $\sim 10^3$  when distance $R$ to the EAS axis changes from $R\simeq 100$ m to $R\simeq 700$ m. It is not what actually observed by the Horizon-T: the density of the particles $\rho(R)$ is not very sensitive to the distance to the EAS axis, and remains essentially flat for the entire region of observations. Furthermore, the magnitude of the density $\rho(R)$ is much higher than it is normally expected for CR energies $\sim (10^{17}-10^{18})$ eV.

\begin{figure}[ht]
	\centering
%	\captionsetup{justification=raggedright}
	\includegraphics[width=0.89\linewidth]{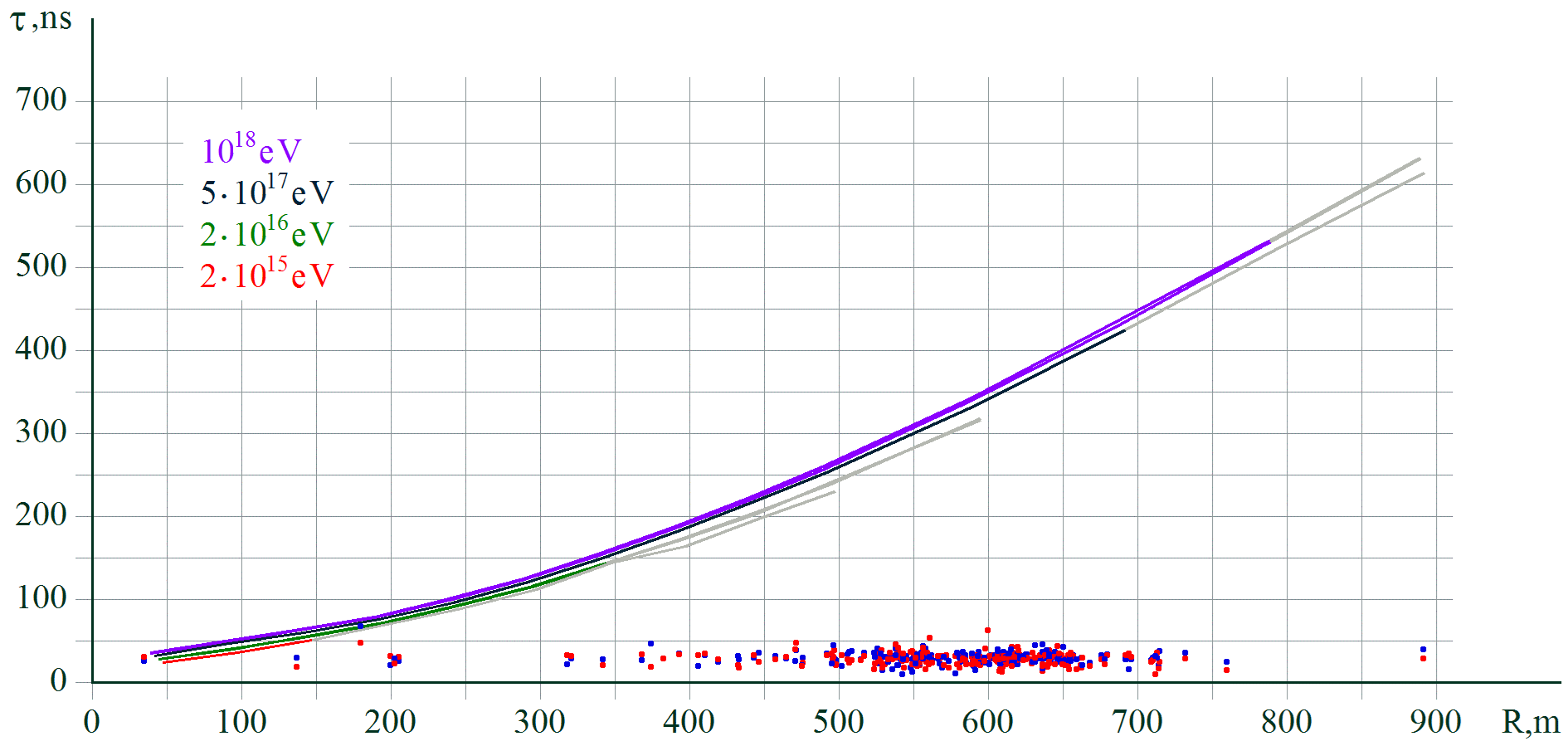}
	\caption{ Solid lines:  simulated EAS disk width versus  distance from axis for different energies shown by different colours.    Pulse width $\tau_{0.8}$ in ns  for the first pulse (in blue) and second pulse (in red) for the bimodal pulses versus  distance to the EAS axis, adopted from \cite{Beznosko:2019cI}.}.  
	\label{width}
\end{figure}

{\bf 3}. The manifestation  of the {\it ``pulse width puzzle"} is  as follows. Fig. \ref{width} presents the CORSIKA-simulated EAS disk width versus  distance from axis for different  energies -- see solid lines in different colours. As explained above,  the thickness of the EAS pancake is supposed to increase with the distance from the core. Therefore, the pulse width must  also increase correspondingly as shown by sold lines on Fig. \ref{width} for different energies. It is not what actually observed by the Horizon-T: all MME events show  similar duration of the pulse width on the level of $(20-35)$ ns irrespective  to  the distance to the AES axis. This observation is in dramatic conflict with conventional picture as outlined above. 

{\bf 4}.   As for the  {\it ``intensity puzzle"}, Fig. \ref{density} suggests that the charged particle  density  $\rho(R)$ varies between $(30-300) $ particles per $m^2$ at the distances $(500-700)$ m from the EAS axis. This is at least factor $10^2$ above the expected  $\rho(R)$ for the primary particle  with energy in the interval  $(10^{17}-10^{18})$ eV. Only the particles  with energies well above $10^{19}$ eV could generate such enormous particle  density as shown on Fig. \ref{density}. However the frequency of appearance  for such highly energetic particles is  only once every few years. Therefore, the intensity of the events estimated from $\rho(R)$  for MME  is several  orders of magnitude higher than the energy estimated by the event rate (on base of standard model of the EASs). This shows  a dramatic inconsistency between the {\it measured} intensity and {\it observed} event rate carried out by  one and the same detector. 

 All the  puzzles presented above imply that the MMEs detected by the Horizon-T are not the conventional CR air showers as they demonstrate enormous inconsistency with standard CR interpretation. What it could be? In the following sections we present arguments that the observed puzzling features could possible be the manifestation of either the dark matter (DM) particles in form of {\it axion quark nuggets} (AQNs) hitting the Earth's atmosphere or signs of propagation of {\it strangelets} through the matter.
  
\section{Possible Physics Explanations of the MMEs: AQN}\label{AQN} 

\subsection{Axion Quark Nuggets (AQN). The basics.} 

 The   AQN dark matter  model  \cite{Zhitnitsky:2002qa} was  invented long ago with a single motivation to   explain in natural way the observed  similarity between the dark matter and the visible densities in the Universe, i.e. $\Omega_{\rm DM}\sim \Omega_{\rm visible}$ without any fitting parameters. We refer to recent brief review  \cite{Zhitnitsky:2021iwg} on the AQN model. Below we outline few key elements of the AQNs of relevance to this proposion.
 
 The AQN construction in many respects is similar to the Witten's quark nuggets, see  \cite{Witten:1984rs,Farhi:1984qu,DeRujula:1984axn}.  This type of DM  is ``cosmologically dark'' not because of the weakness of the AQN interactions, but due to their small cross-section-to-mass ratio, which scales down many observable consequences of an otherwise strongly-interacting DM candidate. 
\exclude{
There are two additional elements in the  AQN model compared to the original models \cite{Witten:1984rs,Farhi:1984qu,DeRujula:1984axn}. First new element is the presence of the  axion domain walls which   are copiously produced  during the  QCD  transition. 
The axion field $\theta (x)$ plays a dual role in this framework: first  it serves  as  an   additional stabilization factor for the nuggets,     which helps to alleviate a number of  problems with the original nugget construction  \cite{Witten:1984rs,Farhi:1984qu,DeRujula:1984axn}.  Secondly, the same axion field $\theta (x)$ generates the strong and coherent $\cal{CP}$ violation in the entire visible Universe at the QCD epoch. }
It  explains the similarity between the dark and visible densities in the Universe, i.e. $\Omega_{\rm DM}\sim \Omega_{\rm visible}$  as as both component, the visible and the dark, are proportional to one and the same dimensional parameter $\Lambda_{\rm QCD}$, see brief review \cite{Zhitnitsky:2021iwg} for the details. 

The key feature of the  AQN model which plays absolutely crucial role for explanations of the MMEs is that nuggets can be made of {\it matter} as well as {\it antimatter} during the QCD transition.  One should emphasize that AQNs are absolutely stable configurations on cosmological scales. Furthermore, the antimatter which is hidden  in form of the very dense nuggets is unavailable for annihilation unless the AQNs hit the stars or the planets when it may lead to observable phenomena. 

 \subsection{AQN-model to explain the MME observations}
   We overview the basic ideas on how the puzzling features listed in Introduction can be understood within AQN framework 
  
   {\it {\bf 1}. ``clustering puzzle":} 
   The multiple number of events is a very generic feature of the system as explained in \cite{Zhitnitsky:2021qhj}.
 Furthermore,  the AQN itself remains almost at the same location as the displacement   $ \Delta l_{\rm AQN}$  during  entire cluster of events is very tiny
   \beq
  \Delta l_{\rm AQN}\sim v_{\rm AQN} \cdot  \tau_{\rm delay}\sim 20~ \rm cm,  ~~~{\rm where} ~~~ \tau_{\rm delay}\sim    \left(10^2-10^3\right) \rm ns
   \eeq
      which implies that  all individual bunches  making  the cluster  are likely to  be emitted by the   AQN along the same direction, and can be recorded  and classified as MME by the Horizon-T. Each event can be viewed as an approximately  uniform front  as mere notion of the ``EAS axis" does not exist in this   framework, see also next item.  However, each individual event may appear to arrive from slightly different direction due to the inherent spread of the emitted electrons at the moment of eruption.  
   
     {\it {\bf 2}. ``particle density puzzle":} Particle density distribution $\rho(R)$  in the AQN framework   shows strong fluctuation from one event to another event.  These variations are mostly related to  the intensity of the individual bursts being  expressed by the number of electrons in the bunch $N$. However, the distinct feature of the distribution in the AQN framework is that it does not depend on $R$,
     as  the notion of the ``EAS axis" does not exist in this   framework as we already mentioned. All these  generic features of the AQN framework are perfectly consistent with the Horizon-T observations as presented on Fig. \ref{density}. However, these observed features   are in  
     dramatic contradiction with   conventional EAS  prediction   shown by solid lines on Fig. \ref{density} with  different colours, depending on energy of the CR. 
          
     Furthermore, the magnitude of the density $\rho(R)$ in the AQN framework (which is mostly determined by parameter $N$)   is   much higher than  one  normally expects for CR energies $\sim (10^{17}-10^{18})$ eV. The corresponding parameter $N$ representing  the number of electrons in the bunch  was not fitted for the present studies to match the observations. Instead, it was extracted from  different experiment in dramatically different circumstances (in proposal \cite{Liang:2021rnv} to explain the ANITA anomalous events as the AQN events). 
     
       {\it {\bf 3}. ``pulse width puzzle":}  In the AQN framework the width of the pulse  $\tau_{\rm pulse}$ cannot    vary much from one event to another. It is a fundamental feature of the framework because the duration of the pulse   is entirely determined by internal dynamics of the AQN during the blast.     
       This feature is in perfect agreement with   observations  \cite{Beznosko:2019cI} 
  \beq
   \tau_{\rm pulse}\approx (20-35) \rm ns ~~~\Leftarrow ~~~[\rm observations ]
  \eeq
  for all recorded MMEs. At the same time, this feature  is in dramatic conflict with conventional picture when the duration of the pulse must depend on the distance to the EAS axis as shown by sold lines  on Fig. \ref{width}. This basic prediction of the conventional CR analysis  is due to increase of the thickness of the EAS pancake with the distance from the EAS axis.   As we already mentioned the mere notions  such as  the ``EAS axis" and the ``thickness of the EAS pancake" do not exist in the AQN   framework.
  
    {\it {\bf 4}. ``intensity puzzle":} Particle density distribution $\rho(R)$  in the AQN framework is estimated in \cite{Zhitnitsky:2021qhj}. The corresponding event to event fluctuations  do not depend on    the distance to the EAS axis as we already mentioned. Such intensity of the events as estimated in \cite{Zhitnitsky:2021qhj} in the AQN framework is consistent with observations shown on Fig. \ref{density}. However, the observations are  in dramatic conflict with conventional CR analysis when such huge intensity could be generated by a primary particle   with energy  well above $10^{19}$ eV with dramatically lower   event rate on the level  of once every few years. The frequency of appearance in the AQN framework as estimated in \cite{Zhitnitsky:2021qhj}    is consistent with observed event rate. 
    
      We conclude this section with the following comment: the emergent picture  suggests  that all the {\it puzzles} formulated above in Section  \ref{confronts} can be naturally understood  within the  AQN framework. One should emphasize  that all phenomenological parameters used in the estimates in \cite{Zhitnitsky:2021qhj} had been fixed long ago  for dramatically different observations in different circumstances for different   environments. 
    
    In this context, it is important to mention that other CR laboratories had recorded a number of unexplained events. In particular, the Telescope Array (TA) collaboration had recorded mysterious bursts (see \cite{Zhitnitsky:2020shd,Liang:2021wjx} for explanations), the Pierre Auger Observatory had recorded the so-called Exotic Events (see \cite{Zhitnitsky:2022swb} for explanations), the ANITA collaboration  observed two anomalous events with non-inverted polarity (see \cite{Liang:2021rnv} for explanations). 
    All these events also present dramatic inconsistencies  with conventional CR picture, similar to our  items {\bf 1-4} discussed above. At the same time all these anomalies can be understood within the same AQN framework with the same set of parameters. What is more important is that the estimations of  the event rates for all aforementioned anomalies (recorded by different instruments: H10T, TA,  AUGER, ANITA) are based on one and the same   flux of the AQNs hitting the Earth. In all those cases the observed   frequency of appearance    of the anomalous events is consistent with the AQN-based estimates, see \cite{Liang:2021rnv,Zhitnitsky:2020shd,Liang:2021wjx,Zhitnitsky:2022swb} for more details. 
  
\section{Stranglets as Possible Physics Explanations of the MMEs}

Ref.\cite{gladysz2021multimodal} offers an alternative explanation of the MME phenomena as a clear manifestation of a strangelet passage through the matter. Such explanation is in line with earlier investigations \cite{asprouli1994interaction} concerning the possible connection between the strongly penetrating cascades and the strangelets suggested to explain  long-range many-maxima cascades observed in the lead chambers of the Pamir and Chacaltaya experiments \cite{buja1981g, pamir1981nuclear}. Similarly, it is hypothesised that the extraterrestial strangelets penetrating deeply into the atmosphere and degradated in the successive interactions with the air nuclei could be responsible for the observed multimodal events. 

      \begin{figure}[ht]
	\centering
	%\captionsetup{justification=raggedright}
	\includegraphics[width=0.49\linewidth]{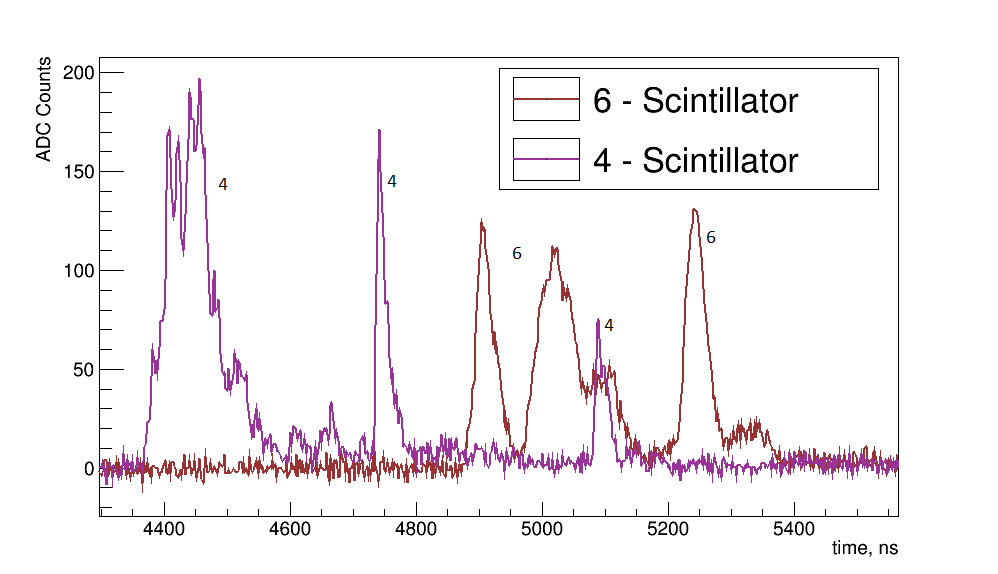}
	\includegraphics[width=0.49\linewidth]{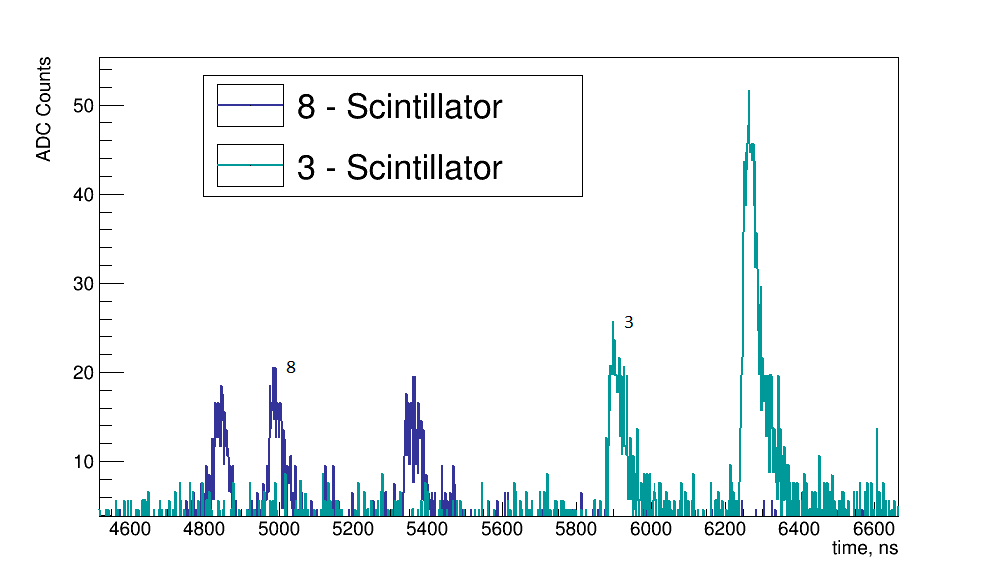}
	\caption{The MME event seen by the HT detector, reprinted from \cite{2019EPJWC.20806002B}. 3 peaks in the station no 8 and 2 peaks in the station no 3 have been detected (right picture). 3 peaks have been detected as well as in the station no 4 and in the station no 6 (left picture). The pulses from these and other detection points demonstrate apparent multimodality. Signals detected by stations located further from the event axis (near point 2) are more delayed as clear correlation between the length of time intervals $\Delta \tau$  and the time arrivals of the successive signals is observed by each station.}  
	\label{TwoMMEs}
\end{figure}

Besides the unusual properties of MMEs discussed above, another interesting feature is the increase of the time intervals between the successive peaks with both the distance of the detection station from the center of the event (event axis) and the time delay of the pulse relative to the first in time pulse registered by this station - see Fig.\ref{TwoMMEs}. Therefore, it can be assumed that the earliest in time peak comes from the highest altitude and possibly produced by the EAS particles born in the first strangelet interaction with the air nuclei. In such scenario, the delay
of the successive peaks and the length of time intervals between them decreases with the increase of the altitude of a possible strangelet interaction. On the other hand the length of the time distances $\Delta \tau$ between the successive modes is increasing with the distance of
the detection point from the EAS axis -- as experimentally observed for the MME events.

\subsection{Multi-maxima MME events as signs of a strangelet passage through the air} 
Strangelets are hypothetical droplets of the stable strange quark matter, see, e.g. \cite{Witten:1984rs}, while "stable" strangelets \cite{gladysz1997strongly} are the long-lived objects capable to reach and pass through the apparatus without decay. Hyperstrange multiquark droplets, having the strange to baryon ratio $f_s$ = 2.2 - 2.6 can be the subject of only the weak leptonic decays and their lifetime is estimated to be longer than $\tau_0 \sim$ 10$^{-4}$ sec. Such long-lived objects should pass without decay not only through, e.g., thick Pb chambers but also, before evaporation into the bundle of neutrons, could travel the long distances in the atmosphere. While traveling, the "stable" strangelet collides and interacts with the matter. The strangelet can be considered as the object of the radius
\begin{equation}
R_{str} = r_0 A^{1/3}_{str} \, ,
\end{equation}
where the rescaled radius
\begin{equation}
r_0 = \Big( {3 \pi \over 2 (1-2\alpha_s/\pi) [\mu^3+(\mu^2-m_s^2)]^{3/2}}\Big)^{1/3} \, ,
\end{equation}
and $\mu$ and $m_s$ are the chemical potential and the mass of the strange quark respectively and $\alpha_s$ is the QCD coupling constant.

\begin{figure}[ht]
	\centering
	%\captionsetup{justification=raggedright}
	\includegraphics[width=0.95\linewidth]{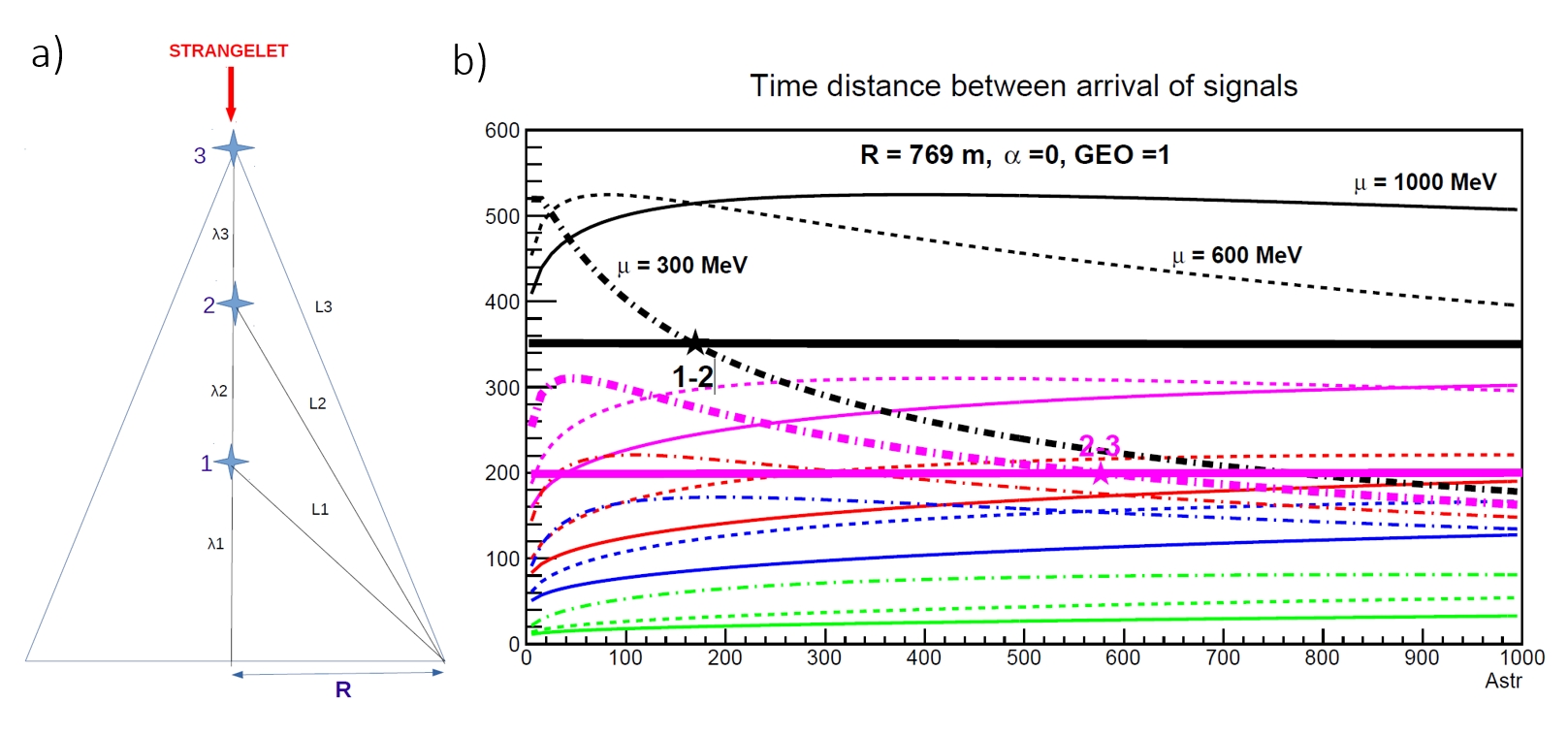}
	\caption{a) left - Scheme of successive strangelet interactions in the air responsible for the appearance of many peak structures in detectors. Only 3 interactions are drawn for illustration. ; b) right - $\Delta \tau$ distances between signals coming from strangelet interactions at the heights of 2$\lambda_{str}$ and 1$\lambda_{str}$ (black), 3$\lambda_{str}$ and 2$\lambda_{str}$ (violet), 4$\lambda_{str}$ and 3$\lambda_{str}$  (red), 5$\lambda_{str}$ and 4$\lambda_{str}$  (blue), 10$\lambda_{str}$ and 9$\lambda_{str}$  (green) above the detector level, as a function of a strangelet mass number $A_{str}$. Calculations have been performed for $\alpha_s$ = 0, $GEO$ = 1, $\mu$ = 300, 600 and 1000 MeV, and $R$ = 769 m from the EAS axis. Experimental $\Delta \tau$ values of Horizon-T MMEs are shown by the straight lines with the same colour as the corresponding $\Delta \tau$ line. Stars indicate intersections of the lines (from \cite{gladysz2021multimodal}).}   
	\label{Strange}
\end{figure}

The idea of a possible strangelet detection by the EAS stations is based on the expectation that strangelets should have geometrical radii much smaller than the ordinary nuclei of the same mass number, and correspondingly, much smaller interaction cross sections. The signals detected by, e.g., the Horizon-T experiment are produced by the nuclear and electromagnetic cascades developing in the air -- light absorber with its density changing with altitude. Fig.\ref{Strange}a illustrates proposed explanation of the multimodal events due to "stable" strangelet interactions in the air.

In such simplified model one assumes that: \\
\begin{itemize}
    \item Each of $n$ signals seen by the Horizon-T detector located at the same distance $R$ from the EAS axis comes from the successive strangelet interaction with the air nuclei. The most delayed peak, marked here by no 1 comes from a strangelet interaction at the altitude $\lambda_1$ above the detection level, peak no 2 from a strangelet interaction at the altitude $\lambda_1+\lambda_2$, peak no 3 from a strangelet interaction at the altitude $\lambda_1+\lambda_2+\lambda_3$, etc.
    \item  The distances between the successive strangelet interactions are equal to the values of the corresponding strangelet interaction paths.
$\lambda_1 \approx \lambda_2 \approx \lambda_3\approx \lambda_n = \lambda =  \lambda_{str}$, where $\lambda$ is the mean value of the strangelet interaction paths for the detected collisions.
 \item The strangelet and the resulting nuclear-electromagnetic cascades 
propagate with the speed of light $c$.
\end{itemize}

In such scenario,  the first detected signal comes from the first strangelet interaction, i.e. the interaction at the highest altitude. The last signal comes from the strangelet interaction at the height $H \approx \lambda_n$ from the detection level. In accordance with the experimental observations, time distances $\Delta \tau$ between the successive signals
decrease with the increase of the strangelet interaction altitude (i.e. $\Delta \tau$ are the longest for the most delay signals, coming from interactions close to the detector level). This effect is additionally enhanced by the increase of the $\lambda_{str}$ value with the
decrease of $A_{str}$ in each of the successive interaction.

Calculations \cite{gladysz2021multimodal} of the proposed scenario are in good agreement with experimental characteristics of the  Horizon-T MME events and predict the observed decrease of $\Delta \tau$ with the increase of $A_{str}$ as well as the increase of the altitude of a possible strangelet interaction - see Fig.\ref{Strange}b. 
\\

{\it To summarize}: Several important features of all MME events are generally in accordance with the strangelet scenario:
\begin{enumerate}
\item The  signals detected by the stations located further from the event axis are more delayed
\item The time intervals  $\Delta\tau$ between the consecutive signals are correlated with their 
arrival times. Most delayed signals have longest  $\Delta\tau$, 
coming from interactions close to the detector level. This effect is aditionally enhanced by the increase
of the strangelet interaction path $\lambda_{str}$ with the decrease of a strangelet mass number $A_{str}$ in each act of the successive interaction.      
\end{enumerate}

Explanations of the {\em "particle density"} and {\em "particle width"} puzzles can be connected with the expected high transverse momentum of the strangelet decay/interaction products and the long distances between the successive strangelet interactions.
  
Authors of \cite{beisembaev2018unusual}  simulated two EAS disks detected in the same distance from the EAS axis but the second disk has been delayed by 100 ns  the first one. In that case {\bf two separate pulses have been detected close to the center of disks}. At longer distances the
multipeak structure was hardly distinguishable. However, the  strangelet interaction paths $\lambda_{str-air}$ are expected to be much longer than  the assumed in \cite{beisembaev2018unusual} 100 ns distance between two EAS disks. Besides that two disks considered in \cite{beisembaev2018unusual} have been assumed to be separated only in time.
The many maxima structure predicted by  the strangelet scenario 
 is the result of as well as a separation in time and the various heights of the successive strangelet interactions.  So, we can expect that in accordance with the observations,  signals from the consecutive strangelet interactions should be separated one to the other also at the longer distances from the event axis.

\section{Future Outlook}

\subsection{Anticipated upgrades of Horizon-T}
Besides the CR-based MME detector complex, there are several other EAS instruments available and operational at the Tien Shan facility and further development of the Horizon-T experiment will make it possible to integrate them as well. For that, it is proposed to place three additional SC-detectors at key points of the complex installation: in the center of a dedicated large area scintillation  EAS shower facility (the "carpet"), in the center of the underground EAS shower detector facility, and above the gamma block of the ionization calorimeter of the "Hadron-55" experiment. Such integration will also require laying new cables, such as for transmitting the master pulse from the center of the Horizon-T installation. Creation of a unified data bank of the upgraded/integrated experimental facility will allow a comprehensive analysis of density and time data on EASs, understanding their relationship, establishing the energy threshold for the appearance of anomalous EAS and their differences from traditional EAS in the energy range above $10^{17}$ eV. That will also make possible detail studies of the longitudinal, spatial, and temporal structure of the detected EAS jets and stems. Installation of acoustic detectors (see Ch.\ref{DAS} below) is also under consideration. 

\subsection{Possible sea level measurements} 

The high importance of this discovery of unusual events calls for the independent verification of the unusual event detection and the accumulation of the data with such events for the subsequent analysis. The plans exist for the development, design, construction, and deployment of DUCK (Detector system of Unusual Cosmic-ray casKades), a cosmic-rays detector system aimed to verify and further study the latest results that were recently presented by the Horizon-T collaboration including the independent verification of the multi-modal event detection and the further study of these events.
The detector will be constructed at the Clayton State University (CSU) campus. The location of the CSU campus is much different from the location of the Horizon-T detector system, the observation level for DUCK is only about 200 m above sea level. 

\begin{figure}[ht]
	\centering
	\includegraphics[width=0.95\linewidth]{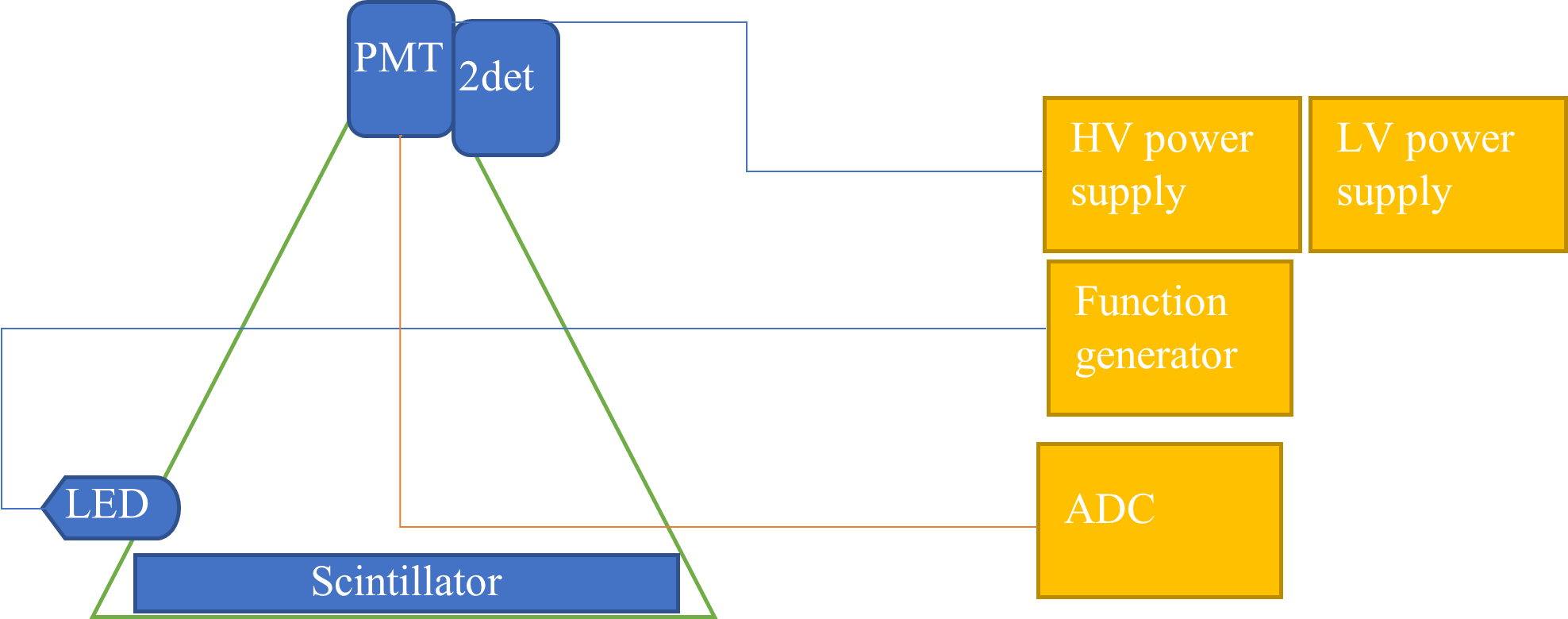}
	\caption{A single scintillator detector module schematic for DUCK system.}   
	\label{DUCK_module}
\end{figure}

The basic detector design for DUCK system is improving upon the original Horizon-T system, and its schematic is shown in Fig. \ref{DUCK_module}. It includes the PMT, the particle detection medium (i.e. scintillator, planned dimensions 1$\times$1$\times$0.1 m) and the LED and secondary detector (2det) for calibration purposes. A green line is the light and weather-tight housing of the module. The detectors are read by the analog-to-digital converter (flash ADC). The detectors are powered by individual mini-module high voltage power supplies (HV in the figure), that connect to common low-voltage power supply (LV). A total for 3 to 4 modules are planned for the initial DUCK system operations and the design is extendable with the funding availability.

\subsection{Radio pulses and the distributed acoustic sensing (DAS)}
\label{DAS}

The key element of the AQN based mechanism explaining  the MMEs is the emission of the large number of relativistic electrons  in form of the bunches. It is known that  
a large number of charged particles $N\approx (10^8-10^9)$ in the background of the geomagnetic field ${\cal{B}}\sim 0.5$ gauss will produce the radio pulse
in both cases: the  CR-induced radio pulse as well as AQN-induced radio pulse  \cite{Liang:2021rnv}.
However, these pulses can be easily discriminated from each other. Indeed, the typical spectral features of the CR-induced radio pulses are: 

1. The generic spectral feature of the CR-induced radio emission is the presence of oscillations which normally start around 100 MHz (depending on the distance from the shower axis). These oscillations are due to the coherence diminishing as the wavelength becomes shorter (in comparison to the ``pancake" size in CR shower).  While it is obviously affected by the detector's filter, this feature is a physical effect due to  changing  number of coherent particles with different wavelengths. 
 
2. Another typical feature of the CR-induced radio emission is that the most of the power is emitted at frequencies around $(20-30)$ MHz for $E_{\rm CR}\approx 10^{17} $\,eV shower. It is a  result of very strong cutoff frequency $\nu_0 < 50$ MHz which strongly depends  on features of the shower.  

3.The cutoff frequency $\nu_0 $
in CR air showers strongly depends on many parameters of the shower such as distance from the central axis when number of particles per unit area strongly depends on this parameter. 
%It must be contrasted with our case of the AQN-induced radio signal when all electrons emitted from the same point at the same instant  are moving along the same direction.  

It should be contrasted with the AQN induced radio pulse with the cutoff frequency $\nu_{\rm c}\sim 0.7$ GHz which is determined by dramatically different physics  \cite{Liang:2021rnv}. 
The main reason for the drastic differences between these two radio pulses is that 
the AQN   event  could be viewed as an (approximately) uniform front of size  several km   with a constant width,   while  EAS  is characterized by central axis.  In different words, the number of particles per unit area $\rho (R)$  in the AQN case   does not depend on the distance from the central axis, in huge contrast  with conventional CR air showers when $\rho (R)$ strongly depends  the distance from the central axis.  The width of the ``pancake" in CR air shower   also  strongly depends on $R$. As a result,   the   effective number of coherent particles contributing to the radio pulse is highly sensitive to the width of the ``pancake"  when it becomes  close to the wavelength of the radio pulse.    These  distinct features lead to very different  spectral properties  of the  radio pulses  in these two cases,  which can be viewed as an independent characteristic of MMEs.   We propose here that  this unique feature  can be used in future  studies for purpose of discrimination and proper selection of the multi-modal clustering events.  

The  mysterious  CR-like  events can also manifest themselves in form of the acoustic and seismic signals, and could be in principle recorded if dedicated instruments are present  in the same area where CR detectors are located. In this case the synchronization between different types of instruments could play a vital role in the discovery of the DM.  In fact, in \cite{Budker:2020mqk}  it has been suggested to use distributed acoustic sensing (DAS) to search  for a signal generated by   an  AQN propagating in the Earth's atmosphere.   It is interesting to note that   a mysterious seismic event 
indeed has been recorded   in infrasound frequency band by Elginfield Infrasound Array  (ELFO). It    has been interpreted in  \cite{Budker:2020mqk}  as the AQN-induced phenomenon.  
 
A complimentary  path to search for  the DM in form of the AQN annihilation events is to study the excess of the radiation in the central regions of the galaxy where DM and visible matter densities are relatively high. In particular,  there is very strong  excess of the  diffuse UV emission which cannot be explained by conventional astrophysical sources   as it dramatically deviates from locations of the UV emitting stars.  As argued in
    \cite{Zhitnitsky:2021wjb} this puzzling diffuse UV emission can be naturally understood  within the same AQN framework. One should emphasize that the corresponding estimates in dramatically  different environment   were based on the same  basic parameters of the AQN model, being used in the present proposal interpreting the MME as the AQN annihilation events in the Earth's atmosphere.

If  our  interpretation of the MME is confirmed by future studies (by studying the synchronized radio pulses or acoustic signals recorded by DAS)
 it would represent  a strong argument  suggesting that  the resolution of two long standing puzzles in cosmology -- the nature of the DM   and the   matter-antimatter asymmetry of  our Universe--  are  intimately  linked. The corresponding deep connection is  automatically implemented 
 in the AQN framework by its construction.

%\section{Acknowledgments}

\addcontentsline{toc}{section}{Bibliography}

\bibliographystyle{atlasnote}
\bibliography{MultiModeWP.bib}

\end{document}